\documentclass{jps-cp}
\usepackage{graphicx}

\title{Internal Plateau in Short GRBs and Quark Stars}

\author{Ang \textsc{LI}}

\inst{Department of Astronomy, Xiamen University, Xiamen, Fujian 361005, China}

\email{liang@xmu.edu.cn}

\recdate{May 27, 2017}

\abst{I summarize our recent calculations on quark stars (QSs), for the purpose of explaining some short gamma-ray bursts characterized by internal plateau in their early X-ray afterglow. According to the present plateau sample, the QS central engine model is demonstrated to more preferred than the original neutron star (NS) one. New QS equation of states (PMQS1, PMQS2, PMQS3) are then proposed, respecting fully the observed burst data and the mass distribution of the Galactic NS-NS systems.}

\kword{quark star, short gamma-ray burst, equation of state}

\begin{document}
\maketitle

\section{Introduction}

Short gamma-ray bursts (SGRBs) are generally believed to originate from the mergers of two neutron stars (NS-NS) or one NS and one black hole (NS-BH). The nature of their central engine remains unknown. The traditional view is that NS-NS mergers produce a BH promptly or less than 1 second after the merger, and accretion of remaining debris into the BH launches a relativistic jet that powers the SGRB. One particular group of SGRBs, however, can not be interpreted in such BH engine scenario, which is characterized by a nearly flat light curve plateau extending to $\sim$ 300 seconds followed by a rapid $t^{-(8 \sim 9)}$ decay~\cite{plateaus}.

A supramassive, strongly-magnetized millisecond NS was proposed to be the candidate central engine of such a SGRB group. From a statistical analysis of a plateau sample, we later find that quark star (QS) remnants might be more preferred than NSs~\cite{ang16prd}. The characterise feature and suggested central engine model are simply depicted in Fig.~1. The present contribution is devoted to the QS central engine model of SGRBs and also provides useful easy-to-use equation of states (EoSs) for these post-merger QSs.

\begin{figure}[h]
\includegraphics[width=16pc]{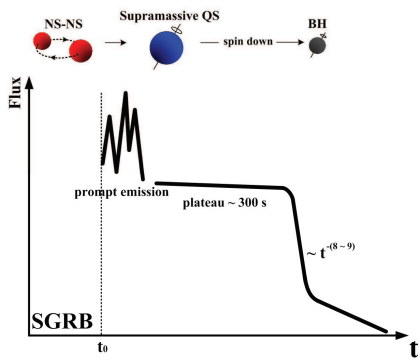}\hspace{2pc}%
\begin{minipage}[b]{18pc}
\caption{Illustration of the characterise feature of SGRBs with plateaus, and the QS central engine model.}\end{minipage}
\end{figure}

In Sect. 2, we first compare theoretical results of the QS central engine model with those of the NS central engine model, and confront them with the observed plateau sample. Then in Sect. 3, for four individual bursts (with known accurate redshift measurements, hence the initial spin period $P_i$ and the
surface dipolar magnetic field $B_p$), we calculate their collapse times (or break time) $t_b$, and further constrain their posterior masses $M_p$ based on various modern NS/QS EoSs. Sect. 4 introduces new supramassive QS EoSs following the finding in the previous section. A short summary is presented in Sect.~5. More details can be found in Refs.~\cite{ang16prd,ang17} and references therein.

\section{QS central engine model}

\begin{figure}[h]
\begin{minipage}{20pc}
\includegraphics[width=20pc]{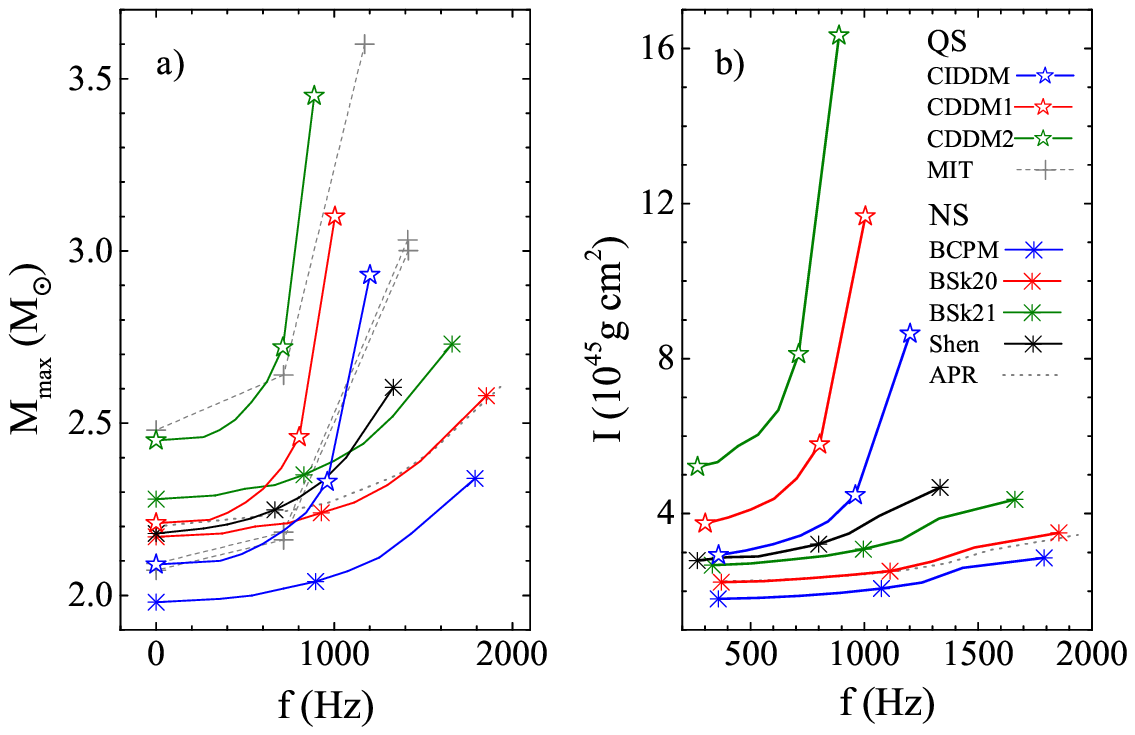}
\caption{Maximum gravitational mass (panel a) and maximum moment of inertia (panel b) as a function of the spin frequency, for three cases of QS EoSs (CIDDM, CDDM1, CDDM2) and four cases of unified NS EoSs (BCPM, BSk20, BSk21, Shen). Previous calculations using the APR NS EoS model~\cite{lasky14} and the MIT QS EoS model~\cite{mit} are also shown for comparison. Taken from Ref.~\cite{ang16prd}.}
\end{minipage}\hspace{2pc}%
\begin{minipage}{15pc}
\includegraphics[width=15pc]{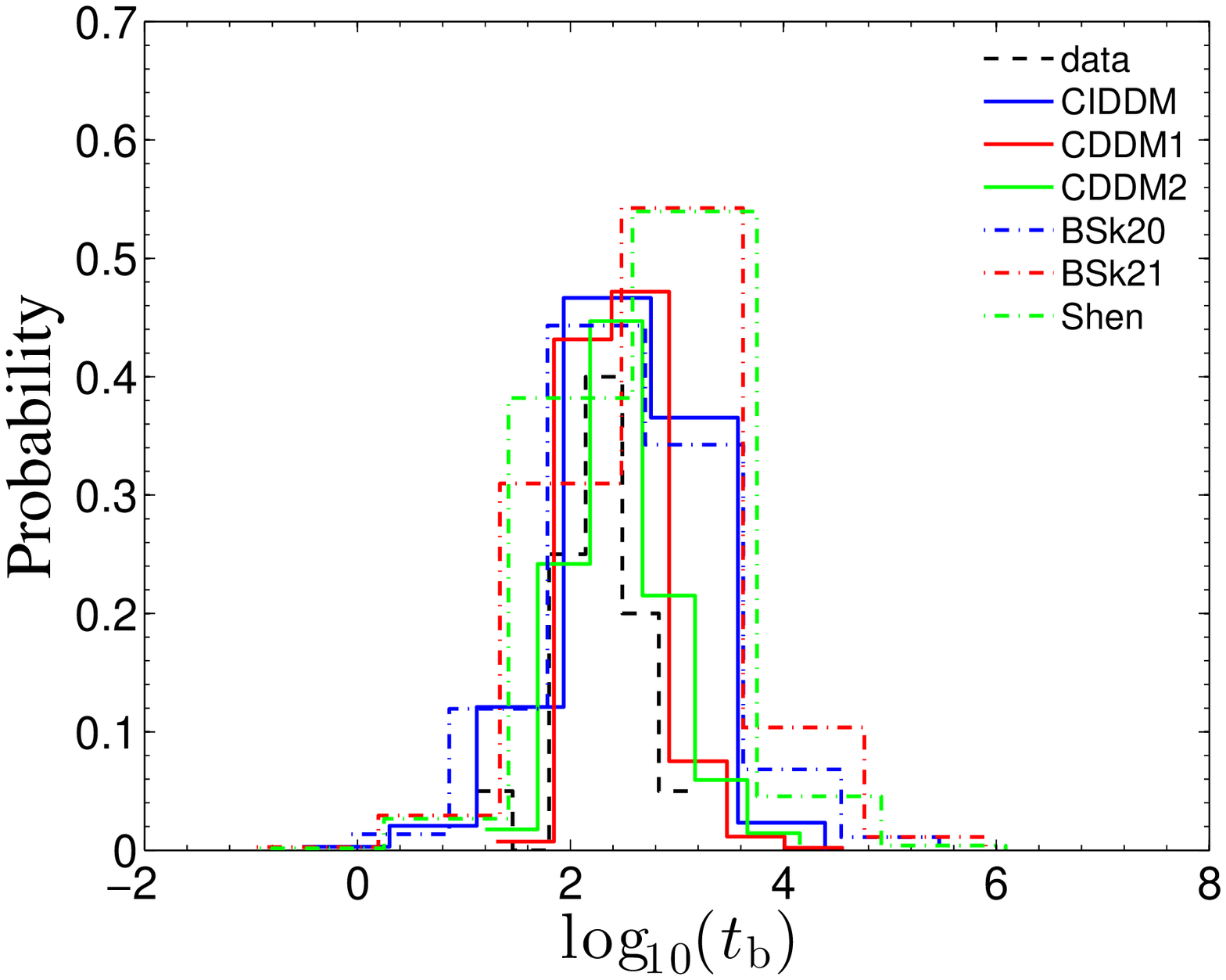}
\caption{Simulated collapse time distributions with three QS EoSs (CIDDM, CDDM1, CDDM2) and three unified NS EoSs (BSk20, BSk21, Shen), as compared with the observed one (dashed curve). Taken from Ref.~\cite{ang16prd}.}
\end{minipage}
\end{figure}

In the QS central engine model (Fig.~1), NS-NS mergers produce a rapidly-spinning, supramassive QS, with the rapid decay phase signify the epoch when the supramassive QS collapses to a BH after the QS spins down due to dipole radiation or gravitational wave radiation. Whether the current modelling of a NS could reproduce reasonably all three observed quantities (the break time $t_b$, the break time luminosity $L_b$ and the total energy in the electromagnetic channel $E_{\rm total}$) is crucially related to the underlying EoS of dense nuclear matter and the calculated rotating configurations.

We first select unified NS EoSs that satisfy up-to-date experimental constraints from both nuclear physics and astrophysics, based on modern nuclear many-body theories, including microscopic methods (e.g.~BCPM) starting from a realistic nucleon-nucleon two-body force (usually accompanied with a nucleonic three body force) and phenomenological ones (e.g.~BSk20, BSk21, Shen), starting from a nuclear effective force with parameters fitted to finite nuclei experiments and/or nuclear saturation properties. Then for QS EoSs, we find typical parameter sets in the new confinement density-dependent mass (CDDM) model, under same constraints of the NS case for high-density part. Three QS EoSs (CIDDM, CDDM1, CDDM2) are employed. The corresponding star properties for all NS/QS EoSs in this study are collected in  Table~I.

For a given EoS, the $rns$ code ($http://www.gravity.phys.uwm.edu/rns/$) presents uniformly rotating, axisymmetric, equatorially symmetric configurations of a NS/QS. In Fig.~2, we perform such calculations up to its Keplerian frequency ($f_{\rm K}$), and the maximum mass $M_{\rm max}$ and the maximum moment of inertia $I_{\rm max}$ are shown as a function of $f$ for both NS and QS EoSs. Evidently, the increases of ($M_{\rm max}, I_{\rm max}$) are more pronounced with the QS EoSs than those with the NS ones. The $M_{\rm max}$ values for the chosen NS (QS) EoSs are roughly $18-20\%$ ($\sim40\%$) higher than the nonrotating maximum mass $M_{\rm TOV}$. The corresponding increase in the corresponding equatorial radius $R_{\rm eq}$ is $31-36\%$ ($57-60\%$). For later use, $M_{\rm max}$ can be fitted well as a function of the spin period ($P$)~\cite{ang16prd,ang17}: $M_{\rm max} = M_{\rm TOV} (1 + \alpha P^{\beta})$. The fitting parameters ($\alpha, \beta$) are also shown in Table I.

By simulating all three distributions ($t_b$, $L_b$, $E_{\rm total}$) of the SGRB internal plateaus sample, we find in Ref.~\cite{ang16prd} that the current modelling of NSs/QSs could qualitatively satisfy the observational constraints of such a SGRB sample, and require the post-merger supermassive stars carrying a strong magnetic field as high as $B_p \sim 10^{15}$ G, an ellipticity $\epsilon$ as low as $0.001$, and an initial spin period $P_i$ commonly close to the Keplerian limit $P_{\rm K}$. In particular, as shown in Fig.~3, the $t_b$ distributions in the QS scenarios are more concentrated, which provide a better agreement with the observed ones. It improves the previous results in the NS EoS cases~\cite{gao16}, without affecting the simulation quantities of other two distributions. Since the selected EoS sample here covers a wide range of maximum mass for both NSs and QSs, there could be some quantitative differences for other EoSs' results, but same conclusions should hold because of the similar NS/QS rotational properties obtained (as compared in Fig.~2). We therefore argue that a supramassive QS is favored than a supramassive NS to serve as the central engine of SGRBs with internal plateaus.

\begin{table*}
\tabcolsep 1pt
\caption{NS/QS EoSs investigated in this study: Unified NS models (BCPM, BSk20, BSk21, Shen), nonunified GM1, APR, CDDM model (CIDDM, CDDM1, CDDM2), and MIT model (MIT2, MIT3, PMQS1, PMQS2, PMQS3). $M_{\rm TOV}$, $R$ are the static gravitational maximum mass and the corresponding radius, respectively; $P_{\rm K}$, $M_{\rm max}$, $I$ are the Keplerian spin limit, the corresponding maximum  mass and maximum moment of inertia, respectively;  $\alpha, \beta$ are the fitting parameters for $M_{\rm max}$. See texts for details.}
\vspace*{-12pt}
\begin{center}
\def\temptablewidth{0.96\textwidth}
{\rule{\temptablewidth}{0.5pt}}
\begin{tabular*}{\temptablewidth}{@{\extracolsep{\fill}}ccc|ccc|cccc|ccc}
   \hline
   & && $M_{\rm TOV}$ & $~R~$ && $P_{\rm K}$ &$M_{\rm max}$& $I$ && $\alpha$ & $\beta$ &Ref.  \\
   &EoS && $(M_{\odot})$ & (km)&& ($10^{-3}$s) & $(M_{\odot})$ & $(10^{45} {\rm g~cm}^2)$ && $(10^{-10}P^{-\beta})$  &&\\
   \hline
   & BCPM   && 1.98 & 9.94  && 0.56 & 2.34& 2.86 && 3.39 & -2.65  & ~\cite{ang16prd} \\
   & BSk20  && 2.17 & 10.17 && 0.54 & 2.58& 3.50 && 3.39 & -2.68  & ~\cite{ang16prd} \\
NS & BSk21  && 2.28 & 11.08 && 0.60 & 2.73& 4.37 && 2.81 & -2.75  & ~\cite{ang16prd} \\
   & Shen   && 2.18 & 12.40 && 0.71 & 2.60& 4.68 && 4.69 & -2.74  & ~\cite{ang16prd} \\
   &  APR   && 2.20 & 10.0  && 0.52 & 2.61& 2.13 && 0.303 & -2.95 & ~\cite{lasky14}\\
   &  GM1   && 2.37 & 12.05 && 0.72 & 2.72& 3.33 && 1.58 & -2.84  & ~\cite{lasky14}\\
   \hline
   & CIDDM && 2.09 & 12.43 && 0.83 & 2.93& 8.645 && 2.58/10$^6$ & -4.93 & ~\cite{ang16prd}\\
   & CDDM1 && 2.21 & 13.99 && 1.00 & 3.10& 11.67 && 3.93/10$^6$ & -5.00 & ~\cite{ang16prd}\\
   & CDDM2 && 2.45 & 15.76 && 1.12 & 3.45& 16.34 && 2.22/10$^6$ & -5.18 & ~\cite{ang16prd}\\
QS & MIT2  && 2.08 & 11.48 && 0.71 & 3.00& 7.881 && 1.67/10$^5$ & -4.58 & ~\cite{ang17} \\
   & MIT3  && 2.48 & 13.71 && 0.85 & 3.58& 13.43 && 3.35/10$^5$ & -4.60 & ~\cite{ang17} \\
   & PMQS1 && 2.31 & 12.75 && 0.79 & 3.34& 10.83 && 4.39/10$^5$ & -4.51 & ~\cite{ang17}\\
   & PMQS2 && 2.46 & 13.61 && 0.84 & 3.58& 13.11 && 5.90/10$^5$ & -4.51 & ~\cite{ang17}\\
   & PMQS3 && 2.59 & 14.33 && 0.88 & 3.75& 15.28 && 9.00/10$^5$ & -4.48 & ~\cite{ang17}\\
      \hline
\end{tabular*}
      {\rule{\temptablewidth}{0.5pt}}
\end{center}
\end{table*}

\section{New QS EoSs}

Since the rapid decay in X-ray luminosity indicates the spindown-induced collapse of a SMS to a BH, one can combine the standard spin-down formula $P(t)/P_0 = \left[ 1 + \frac{4\pi^2B_p^2R^6}{3c^3IP_i^2}t \right]^{1/2}$ with the maximum gravitational mass parameterized as a function of the spin period $M_{\rm max} (P)$, then
setting the protomagnetar mass $M_{\rm p} = M_{\rm max}$, define the collapse time $t_{\rm col}$ as a function of $M_{\rm p}$ for each $P = P_i$:
\begin{eqnarray}
t_{\rm col} = \frac{3c^3I}{4\pi^2B_p^2R^6}\left[ (\frac{M_{\rm p} - M_{\rm TOV}}{\alpha M_{\rm TOV}})^{2/\beta} - P_i^2 \right].
\end{eqnarray}

Based on Eq.~(1), we calculate the collapse time of the selected sample of SGRBs (GRB 060801, GRB 080905A, GRB 070724A, GRB 101219A) with plateaus~\cite{lasky14}. Fig.~4 confronts theoretical results with the corresponding observed values (shown in horizontal lines) for four/five various NS/QS EoSs. The crosses of the EoSs' results with the corresponding horizontal lines in the figure give the predicted protomagnetar masses for the bursts. The input burst data ($P_i, B_p$) with $68\%$ confidence level are listed as well and we employ the most-preferred values for calculations. The shaded region is the $M = 2.46_{\rm -0.15}^{\rm +0.13} M_{\odot}$ mass distribution independently derived from the binary NS mass distribution $M = 1.32_{\rm -0.11}^{\rm +0.11} M_{\odot}$~\cite{2ns}. NS (unified) EoSs are BSk20 and BSk21. NS (nonunified) EoSs are GM1 and APR. QS (MIT) EoSs are MIT2 and MIT3. QS (CDDM) are CIDDM, CDDM1, and CDDM2.

\begin{figure}[h]
\begin{minipage}{18pc}
\includegraphics[width=19pc]{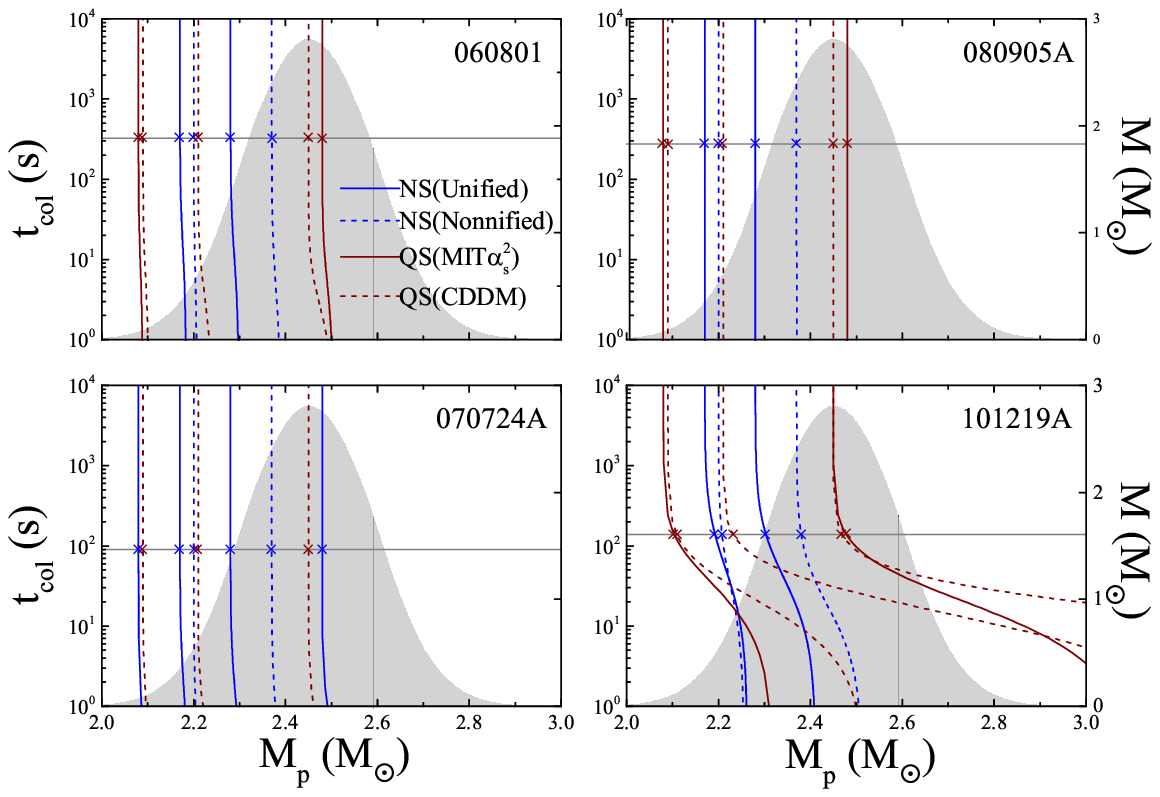}
\caption{Collapse time as a function of the protomagnetar mass for the plateau sample~\cite{lasky14}, to be compared with observed value shown in horizontal lines, respectively. EoSs' details are in Table 1. Taken from Ref.~\cite{ang17}.}
\end{minipage}\hspace{1pc}%
\begin{minipage}{17pc}
\includegraphics[width=18pc]{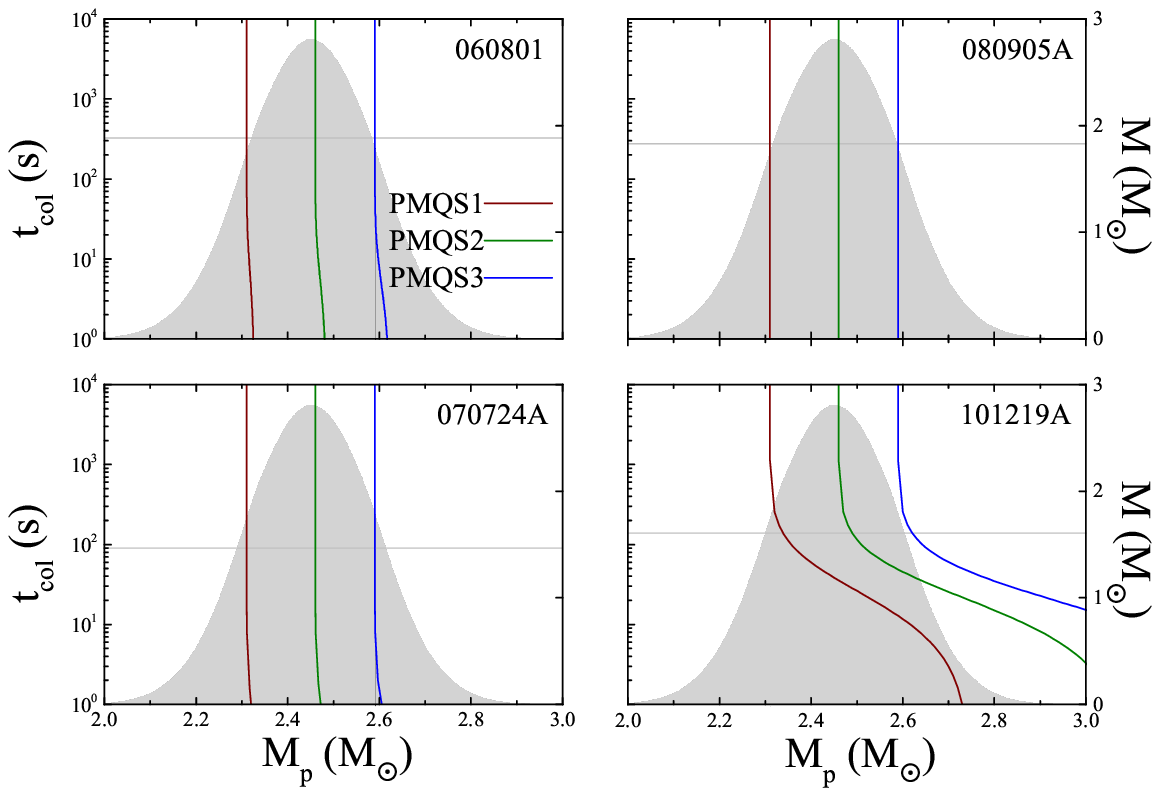}
\caption{Same with Fig.~4, but with three PMQS EoSs newly-constructed in~\cite{ang17}. EoSs' details are in Table 1. Taken from Ref.~\cite{ang17}.}
\end{minipage}
\end{figure}

From Fig.~4, it is clear that the $t_{\rm col}$ vs. $M_{\rm p}$ relations are in all cases here nearly vertical curves before crossing, which demonstrates the required $M_{\rm p}$ related to a burst for each EoS is essentially the EoS's static maximum value: $M_{\rm p} \approx M_{\rm TOV}$. We then argue that how well the underlying EoS reconcile with the current posterior mass distribution (with the mean of the distribution $\sim 2.46 M_{\odot}$), is largely determined by the static maximum mass ($M_{\rm TOV}$) of that EoS. So it is reasonable to construct new compact star EoSs following the observed posterior mass distribution. Within the $95\%$ mass intervals, we then choose three typical values for $M_{\rm TOV}$ of the new EoSs, $2.31 M_{\odot}$, $2.46 M_{\odot}$, $2.59 M_{\odot}$. Considering one of the stiffest unified NS EoS (BSk21) in the market gives $2.28 M_{\odot}$, by reproducing correctly the empirical saturation properties~(see e.g.~\cite{Liapjs,Li2016c}), we restrict ourselves in the present work to new EoSs for only QSs. Moreover, to facilitate a EoS model with as little as model parameters, we employ the MIT bag model for constructing the QS EoSs.

The details of the constructing (named as PMQS1, PMQS2, PMQS3) can be found in Ref.~\cite{ang17}. We also do the $rns$ calculates for the new EoSs and fit their fast-rotating configurations in analytic forms. They are also listed in Table I. The collapse time study can be done for the SGRB sample with these new QS EoSs, shown in Fig.~5. It is clear that these new EoSs' results follow the same conclusion of Fig.~4. Namely, $M_{\rm TOV}$ value of a EoS actually determines its goodness when reconciling with the observed posterior mass distribution. This justifies again the strategy of constructing new EoSs fully respecting the distribution.

To facilitate the usefulness of the new QS EoSs, we have parameterized them in a simple but proper form: $\varepsilon = (9\pi^{2/3}/4)n_{\rm b}^{4/3}/a_4^i + B_ {\rm eff}$, with the parameters shown in details in Table 2 of Ref.~\cite{ang17} for PMQS1, PMQS2, and PMQS3. The pressure is calculated through the standard formulae to fulfill the thermodynamical equilibrium: $P = n_{\rm b}^2(\partial(\varepsilon/n_{\rm b})/\partial n_{\rm b})$. One can find in Ref.~\cite{ang17} for more discussions.

\section{Summary}
In the present proceeding, we study the observed properties of the SGRB internal plateaus sample and reveal the post-merger supramassive stars' physics. QSs are suggested to be the central engine for at least some SGRBs, and NS-NS mergers are a plausible location for quark de-confinement and the formation of QSs.

We proceed to illustrate that how well the underlying EoS would reconcile with the current posterior mass distribution, is largely determined by the static maximum mass of that EoS. We then construct 3 new post-merger QS (PMQS1, PMQS2, PMQS3) EoSs, respecting fully the observed distribution. We also provide easy-to-use parameterizations for both the EoSs and the corresponding maximum gravitational masses of rotating stars. They are welcomed to be used in compact star astronomy, such as SGRB, luminous supernova, GW, kilonova, etc.

\end{document}